# Schottky Collector Bipolar Transistor without Impurity Doped Emitter and Base: Design and Performance

Kanika Nadda and M. Jagadesh Kumar, *Senior Member, IEEE*

*Abstract*— In this paper, we report an alternative approach of implementing a Schottky collector bipolar transistor without doping the ultrathin SOI film. Using different metal work function electrodes, the electrons and holes are induced in an intrinsic silicon film to create the "n" emitter and the "p" base regions, respectively. Using two-dimensional device simulation, the performance of the proposed device has been evaluated. Our results demonstrate that the charge plasma based bipolar transistor with Schottky collector exhibits a high current gain and a better cut-off frequency compared to its conventional counterpart.

## I. INTRODUCTION

Schottky collector bipolar transistors have been studied extensively in literature [1-6]. For example, the vertical Schottky collector PNP transistors on glass substrates have been experimentally demonstrated [3]. The Schottky collector phototransistors can be integrated on silicon waveguides for infrared detection [4]. In hetero-junction bipolar transistors (HBTs), the use of Schottky collector provides a better trade-off between the base-collector capacitance and the collector transit time [6]. In the recent past, lateral Schottky Collector Bipolar Transistors (SCBT) on SOI have been investigated extensively for their less collector storage time and low collector series resistance [7-10] compared to the conventional NPN/PNP BJT. One of the major drawbacks of the lateral SCBT structures is their low collector breakdown voltage. A number of treatments such as an extended BOX region and the two-zone base region have been proposed [10] to increase the collector breakdown voltage of the SCBT. A two-zone base region has a highly doped base at the emitter side and a low doped base region at the Schottky collector metal side. This low doped base region reduces the peak electric field at the metal-base interface, consequently improving the breakdown voltage of the device. However, creating such a two-zone base on ultrathin SOI layers [10] using the conventional doping processes is difficult.

The authors are with the Department of Electrical Engineering, Indian Institute of Technology, New Delhi 110 016, India (e-mail: kanika.shippu@gmail.com; mamidala@ee.iitd.ac.in).

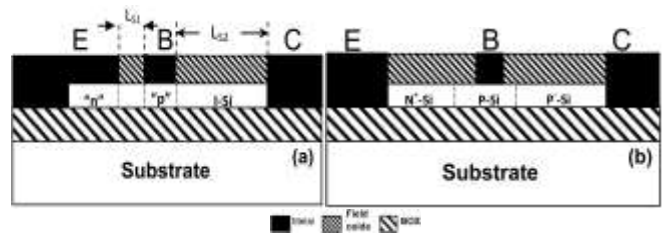

Fig. 1: Schematic cross-sectional view of (a) the proposed SC-BCPT and (b) the conventional two-zone base SCBT.

To overcome the above problem, an innovative Schottky Collector Bipolar Charge Plasma Transistor (SC-BCPT) on undoped silicon is proposed [11-13]. The novel feature of the proposed structure compared to [13] is that it has a Schottky collector and exhibits a larger current gain and $BV_{CEO}\cdot f_T$ product compared to [5]. In the SC-BCPT, the "p" and the "n" regions are formed in the undoped silicon layer by using the emitter metal electrode (work function $\varphi_{m,E} < \varphi_{Si}$) and the base metal electrode (work function $\varphi_{m,B} > \varphi_{Si}$), respectively. The base region consists of an induced "p" region and a lightly doped region of length $L_{s2}$ with $N_A = 1\times10^{14}/\text{cm}^3$. This region $L_{s2}$ on the Schottky metal side reduces the peak electric field at the metal-base interface. Therefore, forming a two-zone base region can be accomplished easily in the SC-BCPT compared to the conventional two-zone base SCBT [10]. Using 2D-simulations, we demonstrate that the SC-BCPT exhibits a significantly higher current gain $\beta$ and cut-off frequency $f_T$ compared to the SCBT with the same device geometry.

## II. DEVICE STRUCTURE AND PARAMETERS

The cross-sectional view of the SC-BCPT and the conventional SCBT are shown in Fig.1. The simulation parameters are given in Table I. In the SC-BCPT, Hafnium metal electrode (work function $\varphi_{m,E}$ =3.9 eV) is used to create the emitter by inducing the electron plasma in the undoped silicon film. Holes are induced to create the base, by using a stack of TiN/HfSiO$_x$/SOI doped with Fluorine (work function $\varphi_{m,B}$= 5.4 eV) [14]. We have chosen the intrinsic silicon film thickness to be within the Debye length [13]. The collector of both the transistors is Aluminium (work function $\varphi_{m,C}$ = 4.28 eV). For the SC-BCPT in Fig 1 (a), the simulated net carrier concentrations (at 2 nm from the top SiO$_2$-Si interface) for zero bias and forward active bias are shown in Fig. 2. Both for thermal equilibrium ($V_{BE} = V_{CE} = 0$ V) and forward active bias ($V_{BE} = 0.7$ V and $V_{CE} = 1.0$ V), the electron and hole concentrations are maintained in the emitter ($n_E = 1×10^{19}$/cm$^3$) and the base ($p_B = 9×10^{18}$/cm$^3$). This creation of the electron/hole concentrations in the undoped silicon film permits the formation of a bipolar charge plasma transistor with a Schottky collector. The conventional two-zone base SCBT with which we have compared our results is also an SOI structure and has the same device parameters as that of the SC-BCPT except that (i) the emitter and the base lengths are chosen to make sure that both the transistors have equal neutral base widths and (ii) the dopings are chosen similar to those given in [10]. Typically, bipolar transistors are compared with identical Gummel numbers. However, in the present case, we could not keep the Gummel numbers of SC-BCPT and SCBT identical. If they are identical, it would make the current gain of the SCBT extremely low (less than 1). The advantage of the SC-BCPT is that the low doped base region towards the collector metal side can be kept as intrinsic.

| TABLE I: Simulation parameters | | |
|---|---|---|
| Parameters | SC- BCPT | Conventional SCBT |
| SOI Thickness (nm) | 30 | 30 |
| BOX Thickness (nm) | 300 | 300 |
| Electrode Sep. (nm) | $L_{s1}$ =10; $L_{s2}$ =200 | $L_{s1}$ =105; $L_{s2}$ =150 |
| Si Film Doping (cm$^{-3}$) | $N_A$=1×10$^{14}$ | $N_A$ =1×10$^{14}$ |
| Emitter conc.(cm$^{-3}$) | "n"=1×10$^{19}$ | $N_D$=5×10$^{19}$ |
| P base conc.(cm$^{-3}$) | "p"=9×10$^{18}$ | $N_A$= 3×10$^{17}$ |
| P$^-$ base conc.(cm$^{-3}$) | $N_A$=1×10$^{14}$ | $N_A$ =1×10$^{16}$ |
| Emitter length (nm) | 50 | 55 |
| Base width (nm) | 40 | P base:145; P$^-$ base:100 |

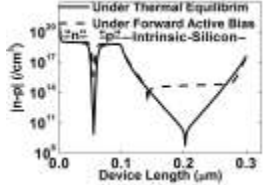

Fig. 2: Simulated net carrier concentrations in the SC-BCPT.

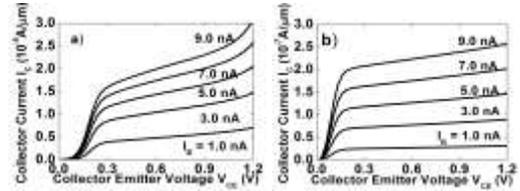

Fig. 3: Output characteristics of (a) the SC-BCPT and (b) the conventional two-zone base SCBT.

Simulations have been performed [15] using the Fermi-Dirac statistics, Philip's unified mobility model [16] and doping-induced band-gap narrowing model [17, 18], all with the default silicon parameters. Our simulation tool does not have an appropriate model for the carrier induced band-gap narrowing. Hence, it is not considered in our study. The standard thermionic emission model [15] is used for the metal contacts of the SC-BCPT. Using the Richardson constant, the surface recombination velocity of electrons and holes are calculated as 2.2×10$^6$ cm/s and 1.6×10$^6$ cm/s, respectively. To account for the Schottky collector junction property, the standard thermionic emission model is used incorporating the effect of image force barrier lowering phenomenon [15]. For recombination, we have enabled Klaassen's model for concentration dependent lifetimes for SRH recombination with intrinsic carrier lifetimes $n_{ie} = n_{ih} = 0.2$ μs [19]. The screening effects in the inversion layer are also considered [20]. The Selberherr impact ionization model is used for calculating BV$_{CEO}$ [21]. The same base metal is used at the base contact of the SC-BCPT and the SCBT to make sure that the base contact properties are identical in both the transistors. This ensures that the lower base current in the SC-BCPT is not due to a difference in the base contact properties. Ohmic contact conditions are assumed at all other contacts with negligible contact resistances.

## III. RESULTS AND DISCUSSION

The output characteristics of the SC-BCPT are compared with that of the conventional two-zone base SCBT in Fig. 3. It is clearly seen that the SC-BCPT device has a higher current driving capability than the conventional SCBT for a given base current. The high V$_{CE}$ offset seen in the output characteristics of the SC-BCPT is due to (i) the large base Gummel number and (ii) the large majority carrier current of the fully depleted and lightly doped p-base-to-metal Schottky diode [3, 7]. The peak current gain of the SC-BCPT (~13000) is several orders higher compared to the conventional SCBT (~40) as shown in Fig. 4. From the Gummel plots shown in Fig. 5, it is observed that the collector current of the SC-BCPT is higher as compared to the SCBT. As we move down the SOI thickness, the built-in-potential of the emitter-base junction of the SC-BCPT reduces (Fig. 6(a)) unlike in the SCBT (Fig. 6(b)). This results in an improved collector current in the SC-BCPT as compared to that in the SCBT for a given forward bias. The Gummel plots in Fig. 5 indicate that the base current of the SC-BCPT is much smaller than that of the SCBT resulting in an enhanced current gain. This reduction in base current is due to the surface electron accumulation at the emitter hafnium-silicon interface [13, 22,

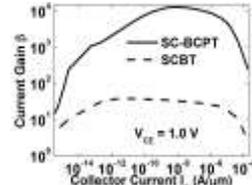

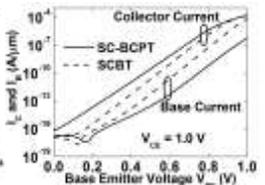

Fig. 4: Current gain of the SC-BCPT and the SCBT.

Fig. 5: Gummel plots of the SC-BCPT and the SCBT.

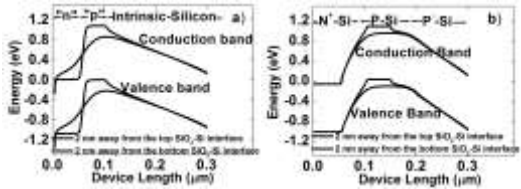

Fig. 6: Energy band diagram of the a) SC-BCPT and b) the conventional two-zone base SCBT.

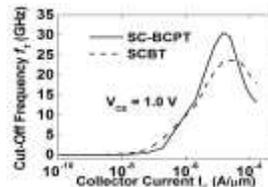

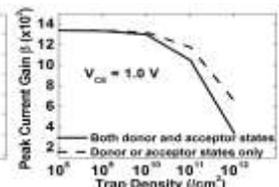

Fig. 7: Cut-off frequency of the SC-BCPT and the SCBT.

Fig. 8: Peak current gain versus trap density for the SC-BCPT.

23]. For $V_{BE} < 0.2$ V, the recombination current dominates as shown in Fig. 5. At higher $V_{BE}$, the diffusion current takes over and the slope of the base current improves. For $V_{BE} > 0.8$ V, the weak slope in the collector current is due to a combination of high injection level effects and the reverse Early effect. In the Gummel plots in Fig. 5, the SC-BCPT exhibits more collector current than the SCBT and therefore, has a higher transconductance leading to a better cut-off frequency $f_T$ (30.56 GHz) than that of the SCBT (23.8 GHz) as shown in Fig. 7. The $BV_{CEO} \cdot f_T$ product of the SC-BCPT is 39.48 V-GHz (@$BV_{CEO}$ =1.3 V) compared to 42.84 V-GHz of the SCBT (@$BV_{CEO}$ = 1.8 V).

## IV. EFFECT OF INTERFACE TRAPS ON CURRENT GAIN

In the SC-BCPT, at the hafnium-silicon interface, the acceptor and the donor type traps could be present [23, 24]. To simulate their influence on the current gain, we have considered both types of traps with the trap energy level (*E.level*) at 0.49 eV from the conduction (or valance) band [24]. The degeneracy factor (*degen*) is 12 and the capture cross-sections for electrons (*sign*) and holes (*sigp*) are $2.85 \times 10^{-15}/cm^2$ and $2.85 \times 10^{-14}/cm^2$, respectively.

The base current increases with an increase in the trap density at the metal-semiconductor interface resulting in a reduction in the peak current gain as shown in Fig. 8. Even with a trap density of $10^{12}/cm^2$, the SC-BCPT still has a substantially higher peak current gain. Controlling the traps at the base contact is also important to realize a large current gain. Prior to the base metal deposition, the surface preparation should be well regulated as is the practise in most advanced fabrication procedures. Inserting a native oxide ~10–15 Å between the metal-semiconductor contact will minimize the effect of surface traps [22, 25].

## V. CONCLUSION

In this paper, for the first time, we have implemented a Schottky collector bipolar transistor with a two-zone base region on an undoped SOI film by inducing charge plasma in the emitter and the base region of the transistor. Device characteristics are simulated and compared with that of a conventionally doped two-zone base Schottky collector bipolar transistor of similar geometry using two-dimensional numerical simulations. Our results demonstrate that the proposed SC-BCPT exhibits a significantly higher current gain, higher cut-off frequency and a better current drive compared to that of its conventional counterpart. However, it is worth noting that the control of metal work function and metal/undoped-silicon contact properties is a challenging issue in the fabrication of the SC-BCPT.

## REFERENCES

[1] F. W. Hewlett, "A compact efficient Schottky collector transistor switch," *IEEE Journ. of Solid-State Circuits*, Vol.14, pp.801-806, 1979.

[2] N. Emeis and H. Beneking, "Fabrication of widegap-emitter Schottky-collector transistor using GaInAs/InP," *Electronics Lett.*, Vol.21, pp.85-85, 1985.

[3] G. Lorito, L. K. Nanver and N. Nenadovic, "Offset voltage of Schottky-collector silicon-on-glass vertical PNP's," *Proc. Bipolar/BiCMOS Circuits and Technology Meeting*, pp.22-25, 2005.

[4] S. Y. Zhu, G.Q. Lo, M. B.Yu, D. L. Kwong, "Low-cost and high-gain silicide Schottky-barrier collector phototransistor integrated on Si waveguide for infrared detection," *Appl. Phys. Lett.*, Vol.93, Issue no.7, Article no.071108, Aug 18, 2008.

[5] K. Nadda and M. J. Kumar, "A novel Doping-less Bipolar Transistor with Schottky Collector," *International Semiconductor Device Research Symposium (ISDRS)*, pp.1-2, 2011.

[6] D. Cohen-Elias, A. Gavrilov, S. Cohen, S. Kraus and D. Ritter, "InP DHBTs Having Lateral and Sidewall Collector Schottky Contacts," *IEEE Trans. Electron Devices*, Vol.60, pp.1163-170, January 2013.

[7] K. Tada, J.L.R. Laraya, "Reduction of the storage time of a transistor using a Schottky-barrier diode," *Proc. IEEE*, vol.55, no.11, pp.2064-2065, Nov. 1967.

[8] M. J. Kumar and D. V. Rao, "Proposal and design of a new SiC-emitter lateral NPM Schottky collector bipolar transistor on SOI for VLSI applications," *Proc. Inst. Elect. Eng.-Circuits, Device Systems*, Vol.151, no.1, pp.63-67, June 2004.

[9] M. J. Kumar and C. L. Reddy, "2D-simulation and analysis of lateral SiC N-emitter SiGe P-base Schottky metal-collector (NPM) HBT on SOI," *Microelectron. Reliab.*, vol. 43, no. 7, pp. 1145–1149, 2003.

[10] M. J. Kumar and S. D. Roy, "A New High Breakdown Voltage Lateral Schottky Collector Bipolar Transistor on SOI: Design and Analysis," *IEEE Trans. Electron Devices*, Vol.52, no.11, pp. 2496-2501, Nov. 2005.

[11] B. Rajasekharan, R. J. E. Hueting, C. Salm, T. van Hemert, R. A. M. Wolters, J. Schmitz, "Fabrication and Characterization of the Charge-Plasma Diode," *IEEE Electron Device Lett.*, Vol.31, pp.528-530, June 2010.

[12] R. J. E. Hueting, B. Rajasekharan, C. Salm, and J. Schmitz, "The charge plasma p-n diode," *IEEE Electron Device Lett.*, vol. 29, pp. 1367-1369, December 2008.

[13] M. J. Kumar and K. Nadda, "Bipolar charge plasma transistor: A Novel Three Terminal Device," *IEEE Trans. Electron Devices*, Vol. 59, No. 4, pp. 962-967, April 2012.

[14] A. Fet, V. Häublein, A. J. Bauer, H. Ryssel, and L. Frey, "Effective work function tuning in high-κ dielectric metal-oxide semiconductor stacks by fluorine and lanthanide doping", *App. Physics Lett.*, Vol. 96, no. 5, Feb. 2010.

[15] *ATLAS Device Simulation Software*, Silvaco Int., Santa Clara, CA, 2010.

[16] D. B. M. Klaassen, "A unified mobility model for device simulation – I: Model equations and concentration dependence", *Solid-state Electron*, Vol. 35, No.7, pp.953-959, July 1992.

[17] D.B.M. Klaassen, J.W. Slotboom, and H.C. De Graaff, "Unified Apparent Band-gap Narrowing in n- and p- type Silicon", *Solid-State Elect.* Vol. 35, No. 2, pp.125-129, 1992.

[18] L. Luo, G. Niu, J.D. Cressler, "Modeling of Bandgap Narrowing for Consistent Simulation of SiGe HBTs Across a Wide Temperature Range," *Proc. Bipolar/BiCMOS Circuits and Technology Meeting*, pp.123-126, Sept. 30-Oct. 2, 2007.

[19] D. B. M. Klaassen, "A unified mobility model for device simulation – II: Temperature dependence of carrier mobility and lifetime", *Solid-state Electron*, Vol. 35, No.7, pp.961-967, July 1992.

[20] M. Shirahata, H. Kusano, N. Kotani, S. Kusanoki, and Y. Akasaka, "A Mobility Model Including the Screening Effect in MOS Inversion Layer", *IEEE Trans. Computer-Aided Design*, Vol. 11, No. 9, pp.1114-1119, Sept. 1992.

[21] S. Selberherr, *Analysis and Simulation of Semiconductor Devices*, Wien, New York: Springer-Verlag, 1984.

[22] M. J. Kumar and V. Parihar, "Surface Accumulation Layer Transistor (SALTran): A New Bipolar Transistor for Enhanced Current Gain and Reduced Hot-Carrier Degradation," *IEEE Trans. Device and Materials Reliability*, Vol. 4, no.3, pp. 509-515, Sep- 2004.

[23] M. J. Kumar and P. Singh, "A Super Beta Bipolar Transistor using SiGe-Base Surface Accumulation Layer Transistor (SALTran) Concept: A Simulation Study," *IEEE Trans. Electron Devices*, Vol.53, pp.577-579, March 2006.

[24] K. Ziegler, "Distinction between donor and acceptor character of surface states in the Si-SiO$_2$ interface," *Appl. Phys. Lett.*, vol. 32, no. 4, pp. 249–251, 1978.

[25] A. Eltoukhy and D. J. Roulston, "The role of the interfacial layer in poly silicon emitter bipolar transistors," *IEEE Trans. Electron Devices*, vol. ED-29, pp. 1862–1869, Dec. 1982.